# GENERAL MARKERS OF CONSCIOUS VISUAL PERCEPTION AND THEIR TIMING

Renate Rutiku[1,2,*], Jaan Aru[2,3], Talis Bachmann[2]

[1] Institute of Psychology, University of Tartu, Tartu, Estonia
[2] Institute of Public Law, University of Tartu, Tallinn, Estonia
[3] Institute of Computer Science, University of Tartu, Tartu, Estonia

*corresponding author: renate.rutiku@gmail.com

## ABSTRACT

Previous studies have observed different onset times for the neural markers of conscious perception. This variability could be attributed to procedural differences between studies. Here we show that the onset times for the markers of conscious perception can strongly vary even within a single study. A heterogeneous stimulus set was presented at threshold contrast. Trials with and without conscious perception were contrasted on 100 balanced subsets of the data. Importantly, the 100 subsets with heterogeneous stimuli did not differ in stimulus content, but only with regard to specific trials used. This approach enabled us to study general markers of conscious perception independent of stimulus content, characterize their onset and its variability within one study. N200 and P300 were the two reliable markers of conscious perception common to all perceived stimuli and absent for all non-perceived stimuli. The estimated mean onset latency for both markers was shortly after 200 ms. However, the onset latency of these markers was associated with considerable variability depending on which subsets of the data were considered. We show that it is first and foremost the amplitude fluctuation in the condition without conscious perception that explains the observed variability in onset latencies of the markers of conscious perception.

# INTRODUCTION

How long does it take from the moment when a stimulus is presented in the environment until the conscious experience of the stimulus starts to arise? Despite the decades-long quest for the neural correlates of consciousness (NCC) it is not known at what time after stimulus onset they occur. Some results suggest that conscious perception is a relatively late process [1,2]. Others point to the importance of mid-latency markers [3]. Still others have found very early correlates for conscious perception [4,5].

One reason for these discrepancies may be that the contrastive method typically used to identify NCC is not only sensitive regarding the actual NCC but also unravels neural processes that precede or follow conscious perception [6]. The contrastive method is supposed to identify markers that are uniquely present or reliably more strongly present in the averaged activity of the condition where a stimulus was consciously perceived compared to the condition where a stimulus was not consciously perceived. However, the markers directly related to conscious perception may not be the only ones that differ between these conditions. Depending on how visual awareness is manipulated and assessed within a given paradigm, neural prerequisites (NCC-pr) and neural consequences (NCC-co) specific to that paradigm may be misclassified as NCC proper [6,7].

Procedural differences between studies can influence the presence as well as the characteristics of the three types of NCC. If experiments employ restricted categories of stimuli it is hard to tell whether the resulting NCC are markers of only one category or whether they can be generalized to other categories as well. For example, the N170 may be a marker of category specific NCC-pr or even NCC proper only for faces [8]. It is also known that the latency of processes correlating with consciousness may shift as much as 100 ms depending on stimulus predictability [9]. If the stimulus set of a study consists of only a few items then perceptual events inevitably become more predictable and the latencies might shift accordingly [9]. Taken together, the fact that a wide variety of different paradigms, stimulus material, recording conditions etc. have been used to study NCC (see [3] for an overview) might at least in part explain why many studies have reported largely different onset times of the signatures of NCC.

However, even if NCC are difficult to compare between studies, can one at least be certain that they are reliable within one study where the paradigm, stimulus material and recording conditions are kept constant? If the same subjects perform the same task over and over again, would contrasting the resulting seen and unseen trials (or representative samples thereof) always lead to comparably similar results in terms of when and where the NCC arise? Looking closer at the rationale behind the contrastive method suggests that this may not be the case. The reliability and onset latency of the markers of the NCC might be affected by a number of different factors even within one study.

For example, it is possible that the latency of the NCC shifts from trial to trial. This would spread out the averaged activity in the condition with conscious perception and the mean onset latency of NCC would become less accurate. A similar effect has been demonstrated for the face-sensitive N170 component if stimulus uncertainty is increased due to added noise [8]. In the worst case scenario latency jitter may even hide the NCC from the contrastive analysis altogether. Results from a contrastive analysis may also be influenced by factors not directly related to the NCC. Different noise profiles may accompany the signal in different trials. Again, this would influence the onset latency of NCC. One assumes that task-irrelevant noise is mostly averaged out when means are created over trials, but this is of course not completely true. Random noise summation will contribute somewhat also to averaged ERPs leading to at least a small effect and thus also on the onset of statistical differences between conditions. This is particularly problematic if the number of trials differs between the contrasted conditions.

To make matters worse, one cannot even be sure that it is only the signal and noise profiles of the condition with conscious perception that dictate NCC reliability and onset latency. The above described concerns apply to the condition without conscious perception as well. This is because for delineating NCC, trials with conscious perception are compared against those without conscious perception of the target. Only the significant differences are considered as candidates of NCC [6], but the reliability and timing of these significant differences also depends on the trials in the condition without conscious perception.

The last consideration is particularly noteworthy in light of the recent work by Schurger et al.[10]. Their results suggest that the pattern of activity in response to unseen stimuli is less stable within and between trials than the pattern of activity in response to seen trials. They used a measure of representational similarity called directional variance. This measure describes how stable the topographic pattern is within a given time window. Note that although directional variance is more sophisticated than the simple ERP calculation the logic behind it is quite similar. It is the core assumption behind ERPs as well that if activity consistently occurs at the same time over trials then it is preserved after averaging whereas inconsistent activity is averaged out. Thus, if directional variance is higher in trials of the unseen condition [10] then it is prudent to assume that ERPs of the unseen condition should also be more variable. Most importantly, this variability will be reflected in the reliability and onset latency of NCC if the contrastive analysis method is used. In other words, trials from the unconscious condition might directly affect the estimated timing of the ERP changes reflecting the NCC.

Taken together, there are several reasons why NCC as identified by a contrastive analysis may vary even within one single study. In order to arrive at a better understanding of the NCC it would be necessary to know how much each of these factors contributes to the results of contrastive analysis. Surprisingly, however, it has not yet been thoroughly characterized how much NCC actually vary when only the data from one experiment are considered.

The present study was designed to address the above described issues. To overcome some of the methodological restrictions of previous studies we employed an experimental paradigm where the role of visual categorical restriction and stimulus predictability were reduced. To that end we used many different stimuli with varying characteristics and we presented these stimuli on perceptual threshold. We hypothesized that for the described paradigm there is at least one marker that distinguishes consciously perceived trials of our heterogeneous visual stimulus set from the non-perceived trials. We call this the general marker of NCC, gmNCC in short. Note, that with "general" we refer to the content-independent nature of the hypothesized gmNCC, because any single stimulus specific NCC-pr, NCC proper and/or NCC-co would not have a critical impact on results if so many different stimuli are considered together.

Our first goal was to investigate which EEG correlates qualify as gmNCC in our experimental paradigm. Our second goal was to study the reliability and any possible variability in the onset latency of gmNCC. Our third goal was to characterize the causes of this variability as thoroughly as possible. To achieve these goals, 100 matched subsets of seen and unseen trials were created by repeatedly sampling from the pool of all available trials. This procedure (depicted in Fig. 2) ensured that objective stimulus content always stayed the same for both conditions while the included trials differed from one matching iteration to another. By performing a contrastive analysis on each of the 100 matched subsets of seen and unseen trials separately and by analyzing variability within these results we show that amplitude variance in the unseen condition has a profound influence on NCC onset latency and sometimes obscures the NCC altogether. Thus, our research may shed light on the question why different studies have found different NCC or report largely different onset times of the NCC.

## MATERIALS AND METHODS

### Subjects

22 subjects participated in the EEG experiment. All subjects were healthy and had normal or corrected-to-normal vision. Data from 4 subjects were not included in the analyses due to a high number of noisy electrodes or too many trials with artifacts. The remaining 18 subjects (8 male) were 18 – 31 years old (mean = 23.2, median = 22, SD = 3.6). 1 subject was left-handed. All subjects gave written informed consent prior to participation and received monetary compensation as a reward. The study was approved by the ethics committee of University of Tartu and the experiment was undertaken in compliance with national legislation and the Declaration of Helsinki.

### Stimuli

The stimulus set consisted of 70 monochrome drawings. The drawings depicted objects from 6 different categories. 4 categories were further divided into line-drawings and solid forms. Thus, there were 10 different types of stimuli: 1. line-drawings of graphical figures, 2. solid graphical figures, 3. short words, 4. line-drawings of man-made objects, 5. solid forms of man-made objects, 6. line-drawings of faces, 7. line-drawings of animated nature, 8. solid forms of animated nature, 9. line-drawings of inanimate nature, 10. solid forms of inanimate nature. Fig. 1 depicts all 70 stimuli sorted by stimulus type. Stimuli were collected from online databases. Occasionally, stimuli were edited manually to keep the number of filled pixels i.e. the contrast energy comparable for all solid forms including text and all line-drawings including faces. There were no important reasons why particular stimulus types or exemplars were chosen. The aim was simply to generate a heterogeneous stimulus set that is comparable to many other related studies. Solid forms were included in addition to line drawings so that both high- and low-frequency information would be presented to the subjects.

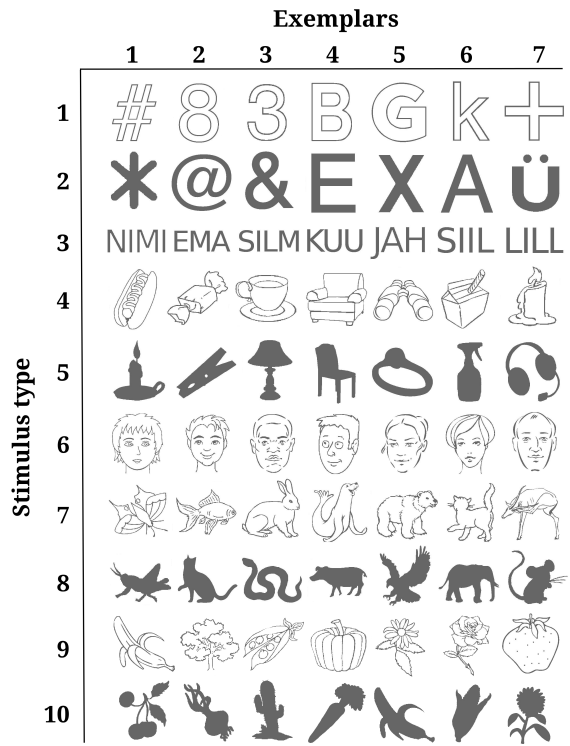

**Fig. 1. All stimuli sorted by type.** Each row depicts all seven exemplars of one stimulus type. Rows are ordered according to the numbers under which the stimulus types are listed in the methods section "Stimuli". The contrast of the stimuli was changed to render them near-threshold, thus making them lighter or darker.

In order to display stimuli at perceptual threshold (i.e. 50% seen responses) their contrast has to be accordingly low. Not all of our stimuli have the same threshold contrast, however. An earlier pilot experiment indicated that for the present stimulus set there are 5 groups of stimuli with roughly similar threshold contrasts within each group: text, solid graphical figures, line-drawings of graphical figures, solid forms of all other figures and line-drawings of all other figures. Thus, contrast was adjusted separately for each of these 5 groups with the help of a short pre-experiment prior to the main experiment (see S1 Text for more details).

Stimuli were presented on a light gray background. Stimulus size was approximately 2.5 degrees of visual angle. Prior to the stimulus a fixation cross was presented (0.35 deg. of visual angle). The response screen contained the question "Did you see something?" in the Estonian language. S2 Text contains more information about the stimulation luminance characteristics.

## Task and design

Subjects were seated in a dark room, 80 cm from the monitor (SUN CM751U; 1024x768 pixels; 100 Hz refresh rate). Each session began with a short pre-experiment to determine the appropriate threshold contrasts for each subject (see S1 Text for more information), followed by the main experiment. The main experiment comprised 770 trials in total. Each of the 70 stimuli was presented 10 times. There were also 70 catch trials where no stimulus was presented. The order of the trials was fully randomized. Each trial began with the presentation of a fixation cross in the middle of the screen for 500 ms. The fixation cross was followed by a blank screen for 750-1250 ms in order to obtain a clean EEG baseline without the ERP of the fixation cross onset or offset. Then the stimulus was presented in the middle of the screen for one refresh frame, i.e. for 10 ms, followed again by the blank screen. After 1s the response screen appeared.

Subjects were instructed to fixate on the cross in the middle of the screen, not to blink until the response screen had appeared, and then to report via button press on a standard keyboard whether they perceived a stimulus on a given trial or not. Seen and unseen responses were given with different hands, but the designated hands were balanced across subjects. There was a break after every 154 trials.

## EEG recording and preprocessing

A Nexstim eXimia EEG-system with 60 carbon electrodes cap (Nexstim Ltd, Helsinki, Finland) was used. All 60 electrodes of the extended 10-20 system were prepared for recording. The reference electrode was placed on the forehead, slightly to the right. The impedance at all electrodes was kept below 15 KΩ. The EEG signal was sampled at 1450 Hz and amplified with a gain of 2000. The bandwidth of the signal was ca. 0.1 – 350 Hz. As our system only allows one pair of eye-electrodes the horizontal electrooculogram (HEOG) was recorded by placing the respective electrodes a few millimeters from the outer canthi of both eyes. Note that blinks could be easily identified in the EEG of posterior scalp sites because the reference electrode was placed on the forehead.

EEG data was preprocessed with Fieldtrip (http://fieldtrip.fcdonders.nl; version 01-01-2013). Trials were epoched around stimulus onset (-500 to +700 ms), re-referenced to the average reference and baseline corrected with a 100 ms time period before stimulus onset. All trials containing artifacts were identified by visual inspection. Trials containing blinks, eye movements, strong muscle activity or other artifacts were completely removed from the data. Noisy signals were interpolated with the nearest neighbor method (see S3 Text for more information on nearest neighbors). 11.6% of trials were rejected due to artifacts on average (median = 10.7%, SD = 6%, range = 4 – 26.9%) and 2.6% of the data was interpolated on average (median = 2.6%, SD = 1.8%, range = 0.1 – 6.3 %). Data were filtered with a 30 Hz low-pass zero phase shift Butterworth filter.

## Data analysis

The behavioral analysis was carried out with the R programming language (http://www.r-project.org/; version 3.1.0). See S4 Text for more information on the behavioral analysis. EEG data was analyzed with Fieldtrip as well as with R.

## Trial matching procedure

In order to find the gmNCC, to study their reliability and any possible variability in their onset latency within one study 100 different matched sets of seen and unseen trials were constructed per subject. The trial matching procedure serves two goals. First, it guarantees that the two contrasted conditions (seen and unseen) are identical with respect to objective stimulus content. Second, it allows us to repeat the contrastive analysis for objectively equivalent matched sets of trials in order to investigate whether the resulting NCC are also equivalent on every iteration.

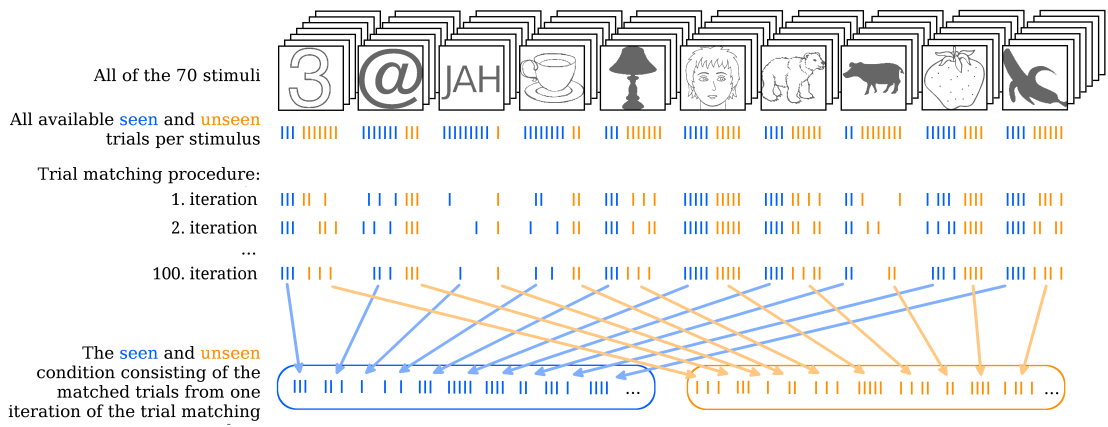

**Fig. 2. Illustration of the trial matching procedure.** The uppermost row indicates all 70 stimuli with one stimulus of each type in the front as examples. Each stimulus was presented 10 times over the course of the experiment. On some of these trials the stimulus was seen, on others it was not. In the second row from above each stroke represents one trial. Seen trials are blue, unseen trials are orange. Note that, for simplicity a total of 10 trials for each stimulus is depicted, but in reality some of these trials were removed during the artifact rejection step of EEG analysis. Therefore, not all stimuli actually had 10 trials left in total. The trial matching procedure would go through all of the 70 stimuli and take the maximal equal amount of seen and unseen trials per stimulus by randomly choosing from the more numerous condition. Three iterations of the total 100 iterations of the trial matching procedure are illustrated as examples. Finally, all the seen and unseen trials that were selected on a given iteration of the trial matching procedure are collapsed into the overall seen and unseen condition and averaged. This step is done for each single iteration, but is here illustrated only for the hundredth iteration as an example.

Fig. 2 illustrates the trial matching procedure. Each of the 100 sets was composed as follows. For every stimulus an equal number of seen and unseen trials were included in the respective conditions. Thus, the algorithm would select a stimulus (the number "3" in the upper left end of Figure 2, for example) and count how many seen and how many unseen trials there are for this stimulus per subject (e.g. 3 vs 7). It would then take all 3 seen trials and randomly choose 3 out of the 7 unseen trials. The algorithm would do the same for all 70 stimuli and pool the chosen trials together into their respective conditions for each subject separately. In case the contrast for one particular stimulus had to be readjusted after the first block of the main experiment (see S1 Text) the algorithm would choose an equal number of seen and unseen trials for each contrast separately.

This random selection of subsets was repeated 100 times for each subject. As a result both the seen and the unseen condition always comprised an equal number of trials for each subject on each iteration of the set matching procedure (m = 122, median = 123, SD = 31, range = 62 to 177). Furthermore, stimulus content also remained identical for both conditions on every iteration. Note that although as a consequence of the trial matching procedure only specific subsets of all available data are considered in the contrastive analyses the amount of trials is still more than typically included or even considered necessary for reliable estimates in ERP research (especially for large components such as the P300; see [11] for a discussion on this topic). After all, even if other studies have used all the recorded trials available to them they are nevertheless also analyzing only a subset of an infinite amount of trials which would maximize the signal-to-noise ratio. Thus the presently employed trial matching procedure should guarantee a sufficiently high signal-to-noise ratio of the experimental conditions and is well comparable to other NCC studies. For example, [1] only had a maximum of 96 trials for the seen and unseen condition together, and that was before artifact rejection. [2] had a maximum of 128 trials for the seen and unseen condition together before artifact rejection. [4] had an estimated average of 100 trials in both seen and unseen conditions. Furthermore, in these studies trial numbers were reported to be only roughly equal between conditions (no further information provided) which may bring its own problems as described in the Introduction.

After the trial matching procedure the seen and unseen conditions comprised 9.6 different types of stimuli on average (median = 10, SD = 0.6, range = 8 – 10), 51.1 different individual stimuli on average (median = 52.5, SD = 8.2, range = 31 – 66) and 1.04 different contrast levels on average (median = 1, SD = 0.1, range = 1 – 1.4). Each individual trial was included in roughly half of the 100 matched sets per person (for seen trials: m = 42%, median = 41%, SD = 14%, range = 19 – 65%; for unseen trials: m = 43%, median = 40%, SD = 18%, range = 18 – 86%). Thus, the matched sets comprised of 42% and 43% of all available seen and unseen trials on average. The mean within-subject difference between the proportion of seen and unseen trials included in the matched sets from all available seen and unseen trials was only 1%.

For comparisons between the unseen and the catch condition all correctly rejected catch trials and the same matched sets of unseen trials were used. The catch condition comprised 59 trials on average (median = 60, SD = 5.6, range = 49 – 68).

## Cluster permutation tests

Differences between conditions were analyzed with nonparametric cluster permutation tests as described in [12] and implemented in Fieldtrip. The advantage of this method is that it identifies significant differences between conditions as clusters evolving over electrodes and time (see Fig. A in S3 Text for an example). Thus it is well suited to study the onset of significant differences without predefining any electrodes or time periods where the effects might occur [13]. After averaging the single trials per condition data points (electrode-time pairs) were compared via dependent samples t-tests. Empirical distributions were created using 10 000 random permutations of the data. The maximal sum of t-values belonging to each cluster was used as the test statistic. Both the entry level for single samples into clusters and the significance threshold for clusters were set at .025. Only clusters lasting longer than 15 ms were considered significant. If not specified otherwise, cluster onsets and offsets were defined as the first/last time points when at least 4 neighboring electrodes showed significant differences between conditions. See S3 Text for more information on neighboring electrodes and the cluster formation.

## Denoising single trials

In order to increase the signal-to-noise ratio for N200 and P300 data were denoised via an algorithm using wavelet decomposition [14]. This method allows the reconstruction of ERP components on the single trial level. The signal is first decomposed into different wavelets and subsequently reconstructed using only those wavelet coefficients that are relevant for the component of interest. Two different sets of wavelet coefficients were used for the reconstruction of the P300 and the N200, but the same sets of coefficients were used for all subjects and all electrodes. All available seen, unseen and catch trials were also always denoised together. For P300 data from electrodes Fcz, C1, Cz, C2, C4, CP1, Cpz, CP2, Pz were denoised. For N200 data from electrodes TP9, TP7, TP10, TP8, P10, P9, O1, Oz, O2, Iz were denoised. These electrodes were selected, because results from seen-unseen comparisons with undenoised data indicated that they constitute the most representative electrodes for N200/P300. More specifically, significant differences between conditions occurred first and lasted longest on these electrodes.

It is important to note that this denoising method can also be applied to data with no clear ERP signal [8,14]. As explained in the introduction and also exemplified in [8] and [14], there are several reasons why event-related signals may not be apparent from averaged data. This method offers one possibility to find out whether any signal may still be present in the single trials or not.

### Correlation tests

To explain the variance in gmNCC onset latencies that remained even after denoising single trial parameters of the two gmNCC (N200 and P300) were extracted from each of the 100 matched sets of trials and correlated with gmNCC onset latencies from the respective contrastive analyses.

First, peak amplitude and peak latency was extracted from the time period of observed variance in the onset latencies of the gmNCC. For each trial, the positive peak between 151 - 268 ms was identified on each of the 9 denoised electrodes belonging to the P300 (Fcz, C1, Cz, C2, C4, CP1, Cpz, CP2, Pz). Similarly, negative peaks were identified between 191 – 232 ms for the 2 denoised electrodes belonging to the N200 (TP7 and P9). These values were averaged per seen and unseen condition for each of the 100 matched sets of trials separately. In addition to mean peak amplitude and mean peak latency, the standard deviation of peak latency was also computed for each matched set. Finally, the 6 parameters (mean peak amplitude, mean peak latency and standard deviation of peak latency for both the seen and the unseen condition) were averaged over electrodes and subjects. Thus, a grand average of all 6 parameters for the N200 and the P300 per matched set was obtained. The grand averages were then correlated with the respective onset latencies of the N200 and the P300 as obtained from the 100 contrastive analyses with denoised data.

In addition to the 12 correlation tests described above 4 confirmatory correlation test were also carried out between averaged ERP parameters and gmNCC onset latencies (see S5 Text for details). All the p-values (n = 16) were corrected for multiple comparisons with the Holm-Bonferroni method.

### RESULTS

### Behavioral results

The false alarm rate in our study was quite low considering the very faint stimulation. The mean percentage of seen reports for catch trials was 4.2% (median = 2.9%, SD = 4.5%, range = 0 – 17 %). Mean detection rate over all stimulus types was close to threshold as intended (m = 51%, median = 48.6%, SD = 13.8 %). The high variance in detection rate stems from the fact that contrasts were estimated separately for different types of stimuli. For several subjects, threshold contrast could not be identified equally well for all stimulus types and detection rates were therefore not always clustered evenly around the mean. S7 Text lists the detection rates for all stimulus types separately and Fig. 3 depicts detection rates for all exemplars within the different stimulus types.

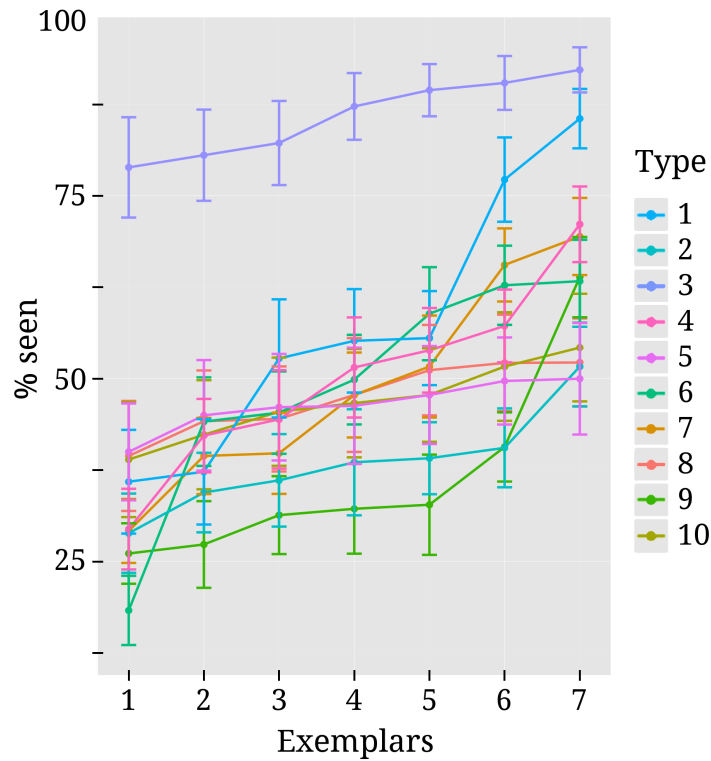

**Fig. 3. Variability in detection rates for exemplars within each stimulus type.** Each colored line corresponds to one of the 10 different stimulus types. They are numbered (on the right-hand side) in the same order as they were listed in the methods section "Stimuli" and depicted as separate rows in Fig. 1. Exemplars 1 to 7 within each stimulus type can also be seen from Fig. 1. Every dot along the x-axis represents one of the 7 exemplars within its corresponding stimulus type. Both here and in Fig. 1 exemplars are ordered according to mean detection rate for convenience of inspection. Vertical lines represent standard errors.

As can be seen from Fig. 3, detection rates are considerably higher for text stimuli compared to other types of stimuli. This was due to the fact that for 12 out of 18 subjects no precise threshold contrast value could be identified for text stimuli. Depending on the contrast, subjects either perceived close to none of the text stimuli or almost all of them. For those subjects the higher contrast level was selected and this pushed the mean detection rate up. For the other nine stimulus types threshold contrasts could be identified more successfully, but there was still variance between individual exemplars. Note, however, that this variability was by and large not systematic across subjects. Most exemplars were perceived above average by some subjects and below average by others.

# EEG markers of conscious visual perception

The first goal of the present study was to identify content-independent general markers of NCC (gmNCC), i.e. markers that distinguish consciously perceived trials of our heterogeneous visual stimulus set from the non-perceived trials. The second goal was to study the reliability and any possible variability in the onset latency of these gmNCC within one study. We therefore did not conduct only one contrastive analysis between the seen and unseen condition, but a 100 of them (see "Cluster permutation tests") in order to compare the results with regard to occurrence and timing of the gmNCC. Importantly, these contrasts were carried out with different matched sets of trials. The matched sets were always identical with respect to the diverse stimulus content, but varied with respect to which specific trials they comprised (see "Trial matching procedure" and Fig. 2).

Results from the 100 contrastive analyses indicate two gmNCC for the present stimulus set - the N200 and the P300. However, results also demonstrate that there is considerable variability in the occurrence and timing of these two gmNCC. Importantly, when we refer to the occurrence and timing of a gmNCC we specifically mean the occurrence and timing of significant differences between the seen and unseen condition. Fig. 4 gives a representative example of the results of one such analysis on one iteration.

The most reliable difference between the seen and unseen condition was the P300. This component was significant in every one of the 100 contrastive analyses and constituted a cluster of 23 electrodes on average (median = 23, SD = 1, range = 21 - 25). The onset latency of the P300 component was not as consistent as its occurrence, however. Fig. 5 contains a histogram of all observed onset latencies of the P300. It is obvious that there are two prominent periods of onset. Mean latency of the first onset period was 143 ms after stimulus presentation (median = 143, SD = 6 ms, range = 128 - 157). Mean latency of the second onset period was 193 ms (median = 190, SD = 13 ms, range = 166 - 223). The P300 was always significant until the end of the tested time period, i.e. 500 ms.

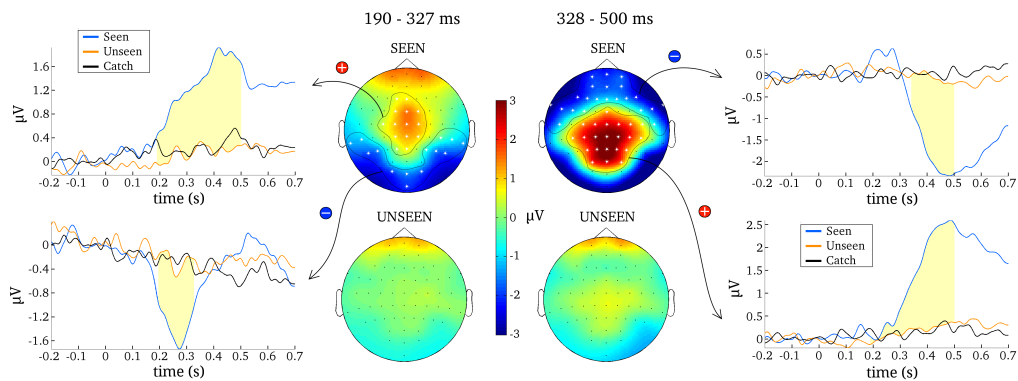

**Fig. 4. Results from one representative contrastive analysis.** Topographies for the seen and the unseen condition are averaged over 190 – 327 ms (left) and 328 – 500 ms (right). ERPs are shown for significant clusters (N200, P300 and late negativity), averaged over all electrodes belonging to each respective cluster (as indicated by white asterisks). Time periods where the seen and the unseen condition are significantly different from each other are colored light yellow.

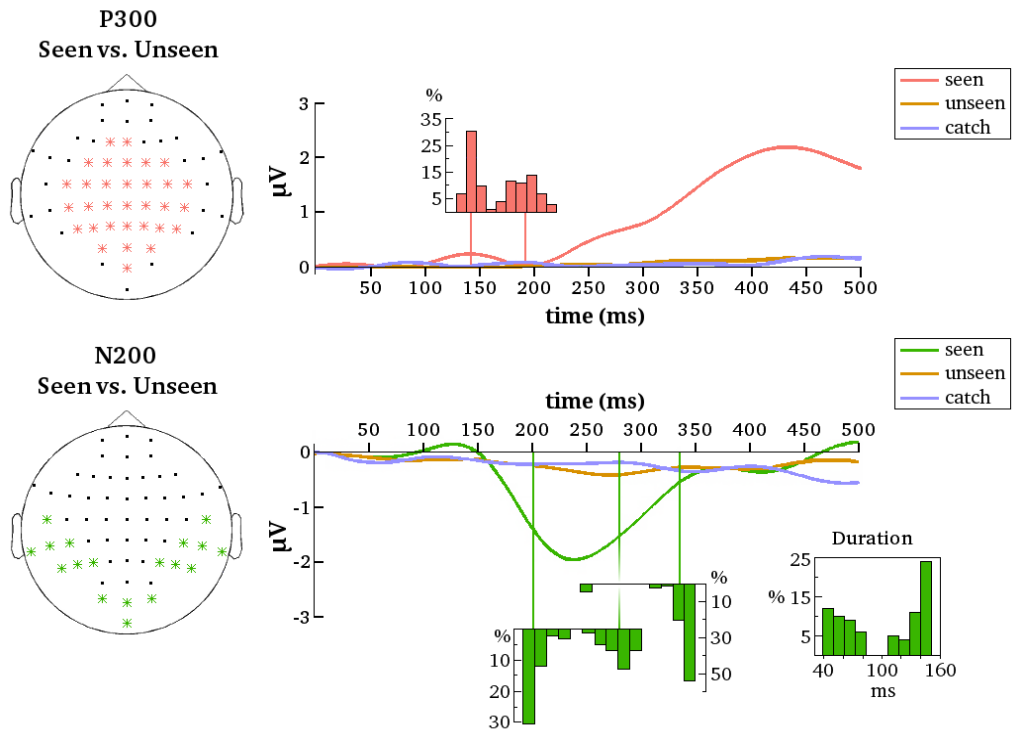

**Fig. 5. Summarized results for all contrastive analyses.** ERPs are averaged over the indicated electrodes (left). These are all the electrodes that belonged to the respective clusters (P300 and N200) for at least 1 of the 100 contrastive analyses between the seen and the unseen condition. Note that because of this averaging not all early differences between conditions – although reliable on several electrodes - may be necessarily apparent from the figure. Histograms depict the distributions of cluster onset times over the 100 contrastive analyses. For N200 there is also a distribution of cluster offset times and of cluster duration. Note that the distributions align with the time axes (in ms).

The N200 was less reliable than the P300. This component was significant in only 81 of the 100 contrastive analyses and constituted a cluster of 10 electrodes on average (median = 11, SD = 3, range = 4 - 15). Thus, in 19% of all cases the contrastive analysis was unable to uncover this gmNCC. Furthermore, even if the N200 was significantly different between the seen and unseen conditions its onset latency nevertheless exhibited considerable variability. As for the P300, there are two prominent periods of onset for the N200. Mean latency of the first onset period was 203 ms (median = 199 ms; SD = 9 ms, range = 192 - 230 ms). Mean latency of the second onset period was 281 ms (median = 281 ms; SD = 12 ms, range = 257 - 301 ms). The onset of the N200 preceded the onset of the P300 in 9% of the contrastive analyses. The duration of the N200 was also divided into two groups. The first group lasted 59 ms on average (median = 57 ms, SD = 12 ms, range = 40 - 82 ms). The second group lasted 137 ms on average (median = 142 ms, SD = 12 ms, range = 108 - 150 ms). The mean offset of statistical significance was at 336 ms (median = 341 ms, SD = 20 ms, range = 245 - 350 ms). Fig. 5 contains histograms of the distributions over all iterations.

Finally, the contrastive analyses also yielded a third component in addition to N200

and P300 which we refer to as the late negativity (see S8 Text for a summary of the respective results). However, the onset latency and topography of this third component suggest that it is probably a consequence of conscious perception [6]. Another alternative explanation is that N200 together with the primary part of P300 constitutes an early effect of conscious perception while the secondary part of P300 and the late negativity constitute a later effect of conscious experience. We leave this problem out of the scope of the present article, however, and will not concentrate on the late negativity any further.

To test if the above described components are uniquely associated with the seen condition we proceeded by comparing the unseen condition to the catch condition. The 100 matched sets of unseen trials were separately contrasted with all available catch trials where the subjects reported not having seen a stimulus. None of the corresponding contrastive analyses yielded any significant differences, however. Thus, it would seem that a condition where the subject did not perceive a stimulus and a condition where there really was no stimulus are indistinguishable in our present dataset at a statistically significant level.

## gmNCC onset variability is partly explained by noise

The above described results suggest that the timing of the two gmNCC (N200 and P300) is highly variable even within one study, ranging over 100 ms depending on which trials are included in the comparisons. In some cases the N200 was even entirely absent. It follows that some variables characterizing single trials are responsible for the varying results and thus the third goal of the present study was to identify these variables. As stated in the introduction, both the signal and the noise profiles of the single trials are potentially involved. It is thus possible that the above described variability in results is not related to the underlying signal profile of the gmNCC at all but stems from irrelevant factors such as an insufficient signal-to-noise ratio or an unequal noise profile between conditions. In order to rule out this possibility data from nine representative electrodes of P300 and ten representative electrodes of N200 was denoised via wavelets (see "Denoising single trials"). Then the 100 contrastive analyses were repeated on the same matched sets of seen and unseen trials as for the undenoised data. Note that the selection of only those electrodes which reliably reflect the N200/P300 components (compared to all 60 electrodes) has no effect on the onset latency of significant differences between conditions because these statistics are conducted point-by-point over time. The results will therefore be comparable with previous results for undenoised data.

After denoising, the onset latency of statistically significant differences again showed considerable variance, albeit with some important differences. The previously observed early period of P300 onsets was effectively not present. Only two from the 100 iterations resulted in P300 onsets earlier than 160 ms. The mean onset latency for the new results was 232 ms (median = 231 ms, SD = 17 ms, range = 151 - 268 ms). Fig. 6 contains the distribution of all onset latencies after denoising the data. Again, the P300 always remained significant until the end of the tested time period and comprised all the 9 electrodes selected for denoising on average (median = 9, SD = 0.1, range = 8 - 9).

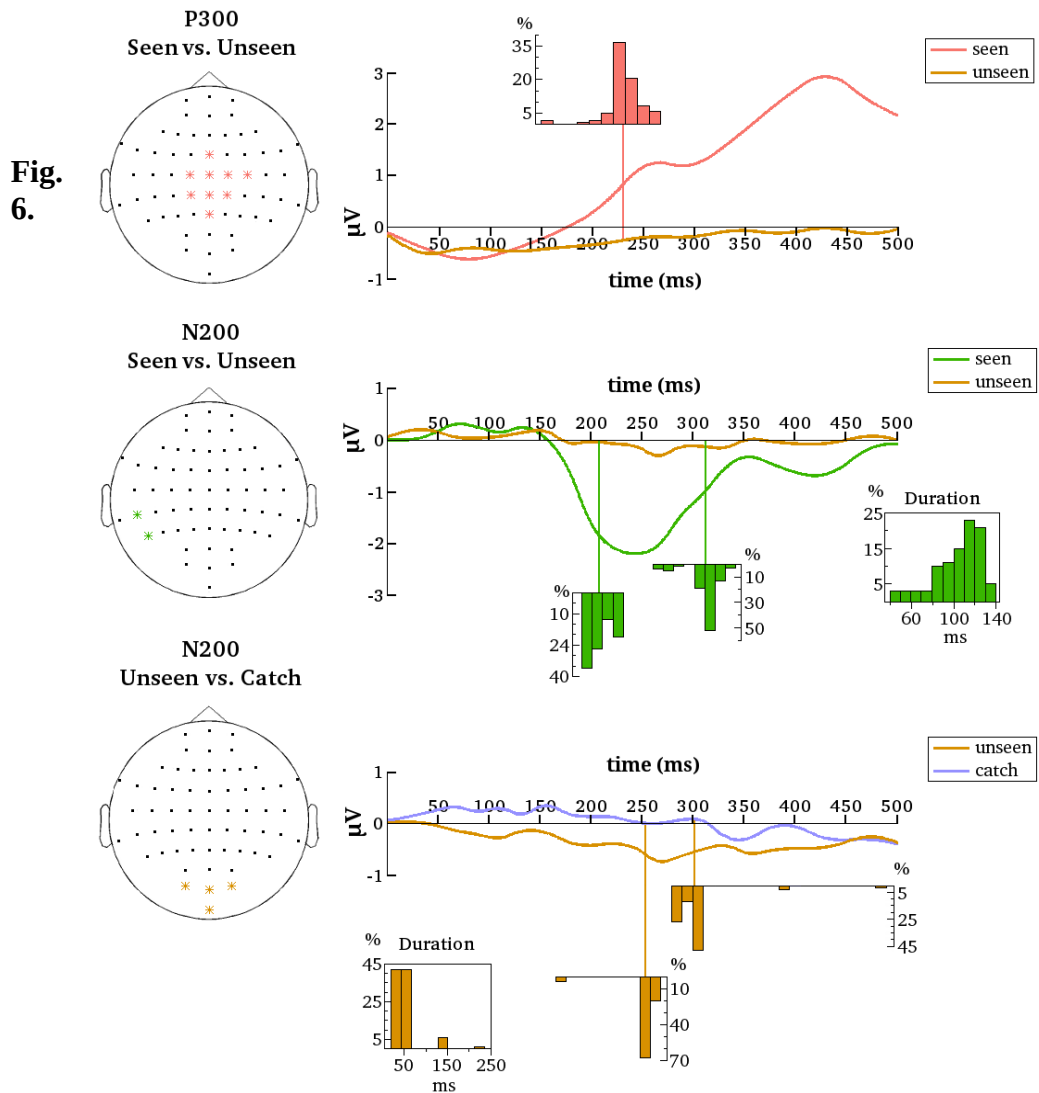

**Fig. 6.** **Summarized results for all contrastive analyses after denoising.** Denoised data is averaged over the indicated electrodes (left). These are all the electrodes that are most representative for the respective clusters (P300 and N200 for the seen-unseen comparisons; N200 for the unseen-catch comparisons). Histograms depict the distributions of gmNCC onset times, offset times and durations over the 100 different contrastive analyses. The distributions align with the time axes (in ms).

Results also changed for the N200. After denoising only 2 temporo-parietal electrodes showed significant differences between conditions. We nonetheless decided to go on with the analyses considering clusters starting from 2 electrodes as significant. For the new results N200 was significant on 97% of the iterations and included 3 electrodes on average (median = 2, SD = 2, max = 9). The mean onset latency was 208 ms (median = 203 ms, SD = 12 ms, range = 191 - 232 ms). The mean offset latency was 313 ms (median = 317 ms, SD = 15 ms, range = 261 - 342 ms). Thus, the mean duration of the N200 was 105 ms (median = 111 ms, SD = 23 ms, range = 39 - 141 ms). Fig. 6 contains histograms of the respective distributions. Note that the N200 onset and duration displayed a highly negative correlation (r = -0.8, t(95) = -12.95, p = 2.2e-16).

To examine if the N200 and the P300 are uniquely associated with the seen condition a separate group of a hundred contrastive analyses comparing the unseen condition to the catch condition were performed. Denoised data were analysed from the same groups of electrodes as for the seen-unseen comparisons. Recall that no corresponding differences for the undenoised data were found, but perhaps the removal of noise will bring to light some subliminal processing of the stimulus in the unseen condition that was previously missed.

As for the undenoised data, there were no significant differences between the unseen and catch conditions on the central P300 electrodes for the denoised data. Thus, the P300 seems indeed to be only associated with the seen condition. The same is not quite true for N200, however. Results revealed a small but quite consistent negative component on an occipital cluster of electrodes. Note that these are not the same electrodes that were most reliable in the seen-unseen comparison. The occipital cluster was significant on 91% of the iterations and included 3 electrodes on average (median = 3, SD = 0.21, max = 4). The mean onset latency of statistical significance was 254 ms (median = 256 ms, SD = 17 ms, range = 166 - 268 ms). The mean offset latency was 302 ms (median = 300 ms, SD = 28 ms, range = 279 - 491 ms). Thus, the mean duration of the occipital negative cluster was 48 ms (median = 43 ms, SD = 33 ms, range = 20 - 235 ms). Fig. 6 contains histograms of the respective distributions.

Results thus suggest that a negative cluster on occipital electrodes can reliably differentiate the unseen condition from the catch condition around 250 ms after stimulus onset. Still, based on consistent differences in topography and latency, one can be fairly confident that it is not the same component as the N200 from the seen-unseen comparison. We therefore conclude that the N200 on a small cluster of left temporo-parietal electrodes is also uniquely associated with the seen condition. This does not necessarily mean that the same neural mechanisms may not be involved in pre-conscious and conscious processing. They can be the same, but the latency of becoming involved and the level of expression of activity are different. Foremost, ERPs are signatures of neural activity rather than neural structure.

### gmNCC onset variability explained by single trial parameters

Results from the previous section indicate that some variability in the gmNCC onset latencies remains even if noise is effectively removed from the data. Thus, some parameters of the gmNCC signal profile must also be involved in the observed variance (see "Introduction" for a description and some theoretical implications of the possible parameters) and it is the third goal of the present study to identify these parameters. Having the list of 100 varying onset times of N200 and P300 one can therefore ask what is different between the matched sets of trials that underlie each of these 100 contrastive analyses.

To answer this question some key parameters of N200 and P300 were extracted from the single trials. Importantly, this was done for the denoised trials, hence effectively subtracting the contribution of noise. More specifically, peak amplitude and peak latency were extracted from the time period of observed variability in onset latencies for both components. Next, grand averages of mean peak amplitude, mean peak latency and mean latency variance were calculated for the seen and the unseen trials and for each of the 100 matched sets separately. Then the 100 different cluster onset latencies for N200/P300 were correlated with these grand averages (see "Correlation tests" for more details).

It is important to note that we are presently not analyzing the peaks of the N200 and the P300 components. Because we are interested in the time period of gmNCC onsets we cannot hope to accurately capture the peaks of the corresponding components in that time window. Our aim is somewhat different. We are trying to understand what happens in the single trials at the time when variance is observed between the 100 contrastive analyses. We are trying to do this by looking at maximal activity in that time window. Because we already have conducted the contrastive analyses, we know that some variables must exist that are responsible for the differences in results. We are now simply taking our analysis one step further by trying to identify these variables.

Fig. 7 illustrates the results of all conducted correlation tests. The onset times of P300 correlated significantly neither with mean peak latency of the seen trials ($r = -0.05$, $t = -0.52$, $p = 1.0$) nor with mean peak latency of the unseen trials ($r = 0.15$, $t = 1.54$, $p = 0.76$). The respective correlations with mean latency variance were also not significant ($r = 0.04$, $t = 0.43$, $p = 1.0$ for seen trials; $r = -0.02$, $t = -0.21$, $p = 1.0$ for unseen trials). There was a moderately significant correlation with mean peak amplitude for seen trials ($r = -0.3$, $t = -3.07$, $p = 0.031$), but the most significant correlation was found with mean peak amplitude for unseen trials ($r = 0.5$, $t = 5.67$, $p = 2.2e-06$).

Results were very similar for N200. The onset times of N200 did not correlate significantly with mean peak latency for the seen nor for the unseen trials ($r = -0.02$, $t = -0.18$, $p = 1.0$ and $r = 0.01$, $t = 0.09$, $p = 1.0$, respectively). The correlations with mean latency variance were also not significant ($r = 0.22$, $t = 2.25$, $p = 0.21$ for seen trials; $r = -0.23$, $t = -2.26$, $p = 0.21$ for unseen trials). The correlations with mean peak amplitudes of the seen and the unseen trials were again significant ($r = 0.34$, $t = 3.49$, $p = 0.009$ and $r = -0.48$, $t = -5.4$, $p = 7e-06$, respectively).

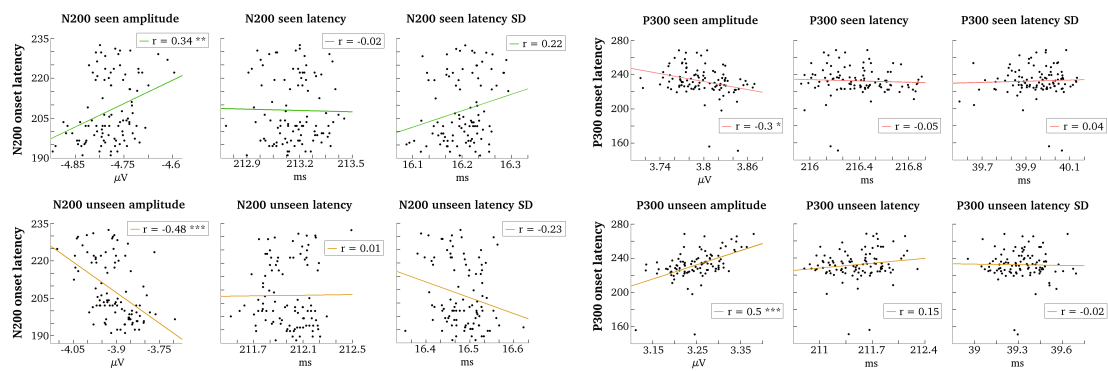

**Fig. 7. Correlations between gmNCC onset times and single trial parameters.** Grand averages of single trial N200/P300 parameters (amplitude, latency and standard deviation of latency) were correlated with N200/P300 onset times (indicated in ms on the y-axes). Correlation tests are carried out separately for seen and unseen trials. P-values < 0.05 are indicated with *. P-values < 0.01 are indicated with **. P-values < 0.001 are indicated with ***.

To exclude any possible confounds with latency variance and to demonstrate more convincingly the relevance of the amplitude parameter for the observed variability in gmNCC onset times, the above analysis was repeated by first averaging single trials and then extracting peak amplitude. The results are presented in S6 Text.

Finally, to be sure that the above described results are meaningful and do not derive from the simple fact that any activity in the unseen condition – if at all present - is much weaker compared to the seen condition we repeated all the contrastive analyses and correlation tests, but replaced the post-stimulus time window of the unseen condition with baseline data. Results are described in S9 Text. These analyses show that both for the P300 and N200 variability in onset latencies is much decreased compared to the results presented above and no significant correlations with mean peak amplitude of the "unseen" condition (i.e. baseline activity) remain.

We conclude that besides noise the varying onset times of the two gmNCC are first and foremost explained by amplitude variability in the unseen trials, but amplitude variability in the seen trials has an effect as well. If the range of mean peak amplitude values for the seen and the unseen trials in Fig. 7 are compared, it can be noticed that mean peak amplitude of the unseen trials varies over a wider range than mean peak amplitude of the seen trials. Thus, it is not surprising that this variability is reflected in the onset times of significant differences between the seen and the unseen condition.

Importantly, there is no evident connection between gmNCC onset times and latency parameters. And indeed, if one takes a look at the distributions of mean peak P300 and mean peak N200 latencies in Fig. 7, one can observe that the variability is very small in absolute numbers. It seems that the mean peak latencies of the two gmNCC are very similar across the different matched sets of trials. The distributions of mean latency variance for P300 and N200 in Fig. 7 make it clear that peak latency shifts considerably over single trials, but mean latency variance is again very similar across the 100 different sets of trials.

## General vs. specific markers of NCC

Although the aim of this study was to find and describe content-independent general markers of conscious perception it must be noted that not all NCC have to be general. There might exist specific markers of conscious perception which are associated with certain stimulus types only (e.g. N170 for faces [8]). Our rationale was to capitalize on a heterogeneous stimulus set so that no stimulus specific markers (whether NCC-pr, NCC proper and/or NCC-co) could dominate the results. Thus, we presently did not aim to differentiate between general and specific markers of conscious perception nor to investigate them parametrically. These questions will have to be addressed in future research.

On the other hand, even if a marker is in essence the same for different stimulus types (I.e. it marks the same underlying neural process) its latency and/or amplitude may probably still vary due to stimulus characteristics or perceptual quality. We have conducted some preliminary analysis in this regard as far as the dataset allows. Comparisons between different subgroups of stimulus types and between stimuli with higher or lower detection rates are presented in S10 Text. These results do not indicate any influence of stimulus characteristics on N200 and P300 amplitude/latency. However, differences in detection rate seem to be associated with systematic amplitude and/or latency modulations for both N200 and P300.

# GENERAL DISCUSSION

The first goal of the present experiment was to find general markers of NCC (gmNCC), that is - markers that distinguish consciously perceived trials of a heterogenous visual stimulus set from the non-perceived trials. The second goal was to study how much these gmNCC vary within one experiment. The third goal was to characterize the causes of this variability as thoroughly as possible. A heterogeneous visual stimulus set was presented at a near-threshold contrast. Thus, our paradigm was designed to reduce the influence of stimulus predictability and categorical specificity. 100 different matched subsets of the resulting seen and unseen trials were contrasted to identify the gmNCC, to study their reliability and variability of their timing. Results indicate that N200 and P300 are the two gmNCC for our paradigm, but their onset latency exhibits considerable variability.

## Generality of various NCC

Different onset latencies of NCC observed in different studies have previously been explained with differences in stimulus material and tasks [15]. One explanation is that depending on how visual awareness is manipulated and assessed within a given paradigm, neural prerequisites (NCC-pr) and neural consequences (NCC-co) specific to that paradigm may be misclassified as NCC proper when the contrastive method is used [6,7]. We were able to show, however, that NCC can vary even within one study where the paradigm, stimulus material and recording conditions were kept constant. Admittedly, the present paradigm is also not sufficiently free from possible confounding factors so as to confidently argue that N200 and P300 really are the NCC proper. We can only argue that for our study these ERP components which may be markers of any one of the 3 subtypes of NCC are general enough so that they do not emerge as related to some narrow visual categorical stimulus group. For this reason we call them general markers of NCC. The problem is simply that besides general NCC proper there might also exist general NCC-pr or general NCC-co. On the other hand, even with regard to the NCC proper we should not think that conscious experience marked by it must be invariant and narrowly fixed in time. Conscious experience of the target stimulus need not be indicated by a certain type of strictly defined NCC, but could be understood as a successful evolution of necessarily required neural events over time (see [8, 10, 16] for similar arguments).

The P300 component is a well known marker of conscious perception. It has been found in almost all electrophysiological studies investigating the ERP-correlates of consciousness. Only when the same experimental stimuli are presented repeatedly [17,18] or when one has prior knowledge about the presented stimulus [9] does the P300 increment does not occur as a difference between trials with and without conscious perception. As P300 might reflect updating of working memory (WM) [19], which is arguably not needed when the very same stimuli are already encoded in WM, P300 is not a marker of conscious perception under such experimental conditions [9]. For the present study stimuli were deliberately unpredictable. Thus, in light of the argumentation presented above it is possible that the P300 is not a gmNCC proper, but rather reflects a general process following the NCC proper, i.e. it represents the NCCco.

The N200 has also been found as a marker of conscious perception, but not as often as the P300. In many studies the N200 is not reliably different between conditions with and without conscious awareness [1,2]. The present results offer an explanation for these varying results. As the reliability of this marker of conscious perception depends on which single trials are included in the seen as well as the unseen condition it is possible that previous studies have simply missed it. This possibility has also been noted by [2]. Nonetheless, the present results are different because there is no clear N200 component present in the unseen condition. Studies using stronger stimuli find a well pronounced N200 which is not different between conditions [1,2]. Thus, one might argue that the N200 reflects a general process preceding the NCC proper.

Yet, it seems for the undenoised data that the average onset of P300 occurs somewhat earlier than the average onset of N200. This would be in conflict with the view that N200 reflects a pre-conscious process prior to the NCC proper or NCC-co, which is P300. Another interesting observation is that both components show two periods of onset for the undenoised data. One explanation for these results is that the abnormally distributed results are due to a confounding signal in the measurements (e.g. alpha oscillations) and are actually not a property of the gmNCC per se. The current results favor this explanation because after denoising the relevant single trial data, the discrepant periods of onset disappear. After denoising both components are still reliably associated with conscious perception, but they show one fairly similar period of onset which falls around 200 ms after stimulus presentation. Thus, noise seems to explain a big part of the initial variability in gmNCC onset latencies and the extremely early onset latencies of the P300 in particular.

Despite the fact that the very early period of P300 onsets disappeared after denoising the EEG signal it is noteworthy that P300 still sets on somewhat earlier than is typically estimated in other relevant studies (around 270 ms in [2], for example). One explanation for this discrepancy may be that we are presently not capturing specifically the onset latency of the P3b subcomponent which is arguably the most relevant P300 subcomponent for conscious perception [15]. P300 also has a somewhat earlier subcomponent – the P3a. It is evident on fronto-central electrodes and is hypothesized to reflect automatic and possibly nonconscious orienting responses (e.g. 20]). Perhaps in our study a stronger P3a response occurs for the consciously perceived stimuli and this is the earliest critical difference within the P300 that we capture with our contrastive analyses. In that case the earliest part of P300 may just as well reflect a general process of gnNCCpr preceding the NCC proper for our paradigm.

## gmNCC onset variability explained

After noise was removed from the data we were able to show that variance in the gmNCC onset latency could be first and foremost attributed to amplitude variance in the unseen condition. Amplitude variance in the seen condition was also associated with the varying gmNCC onsets, albeit to a lesser extent. It is important to note, however, that not only were there no clear N200 and P300 components in the unseen condition, but there really were no clearly pronounced ERP components associated with the unseen condition at all (see [21] for similar results). Thus, the question arises whether this fact in itself could explain the results showing that most of the variance in gmNCC onset latencies came from the unseen condition. To test this possibility we repeated all the analyses after replacing the post-stimulus data of the unseen condition with baseline data. This lead to a marked decrease of variability in gmNCC onset latencies compared to results with actual data and no significant correlations with the amplitude of the unseen condition (i.e. baseline activity) remained. This fact speaks against the possible confound of an unequal signal-to-noise ratio between the seen and the unseen condition in the present study. Furthermore, despite the lack of any clear ERP components in the unseen condition it still exhibited reliable differences with respect to the catch condition on occipital electrodes around 250 ms after stimulus presentation – supporting the assumption that there is a weak signal and thus a weak ERP in the unseen condition. The activity may just be too weak to form a clear component on the ERP.

Although the same occipital electrodes that differentiated the unseen condition from the catch condition sometimes also showed significant differences between the seen and the unseen condition, these were not the most reliable electrodes for the N200 of conscious visual perception. N200 was most reliable on left temporo-parietal electrodes in the present study. Thus, one additional possibility why some previous works have not found the N200 as a marker of conscious perception could be because it is mixed up with other posteriorly recorded components that have similar latencies, but are not necessarily associated with conscious perception.

Taken together, the results reported in this study suggest that signal properties of the unseen condition (amplitude fluctuations in particular) can have a noteworthy impact on the results of a contrastive analysis. Although such effects are generally expected their extent has not been thoroughly investigated in previous studies. However, the present study is comparable to another recent study [10]. The authors of this study elegantly showed that the pattern of activity in response to unseen stimuli is less stable within and between trials than the pattern of activity in response to seen trials. Thus, instability may be a property of unconscious neural responses while stability constitutes a hallmark of conscious perception. Our results confirm this assumption, but in addition we show that because of this difference in stability comparisons between the seen and unseen condition can yield widely varying results in terms of when and where significant differences begin to occur.

## Theoretical implications

The P300 as a marker of consciousness is most consistent with the theory of a global workspace consisting of multiple areas including frontal, parietal, and temporal cortices [22,23]. We cannot say anything certain about the sources of our P300, but since it is a well-studied component one can be fairly confident that a similar multi-focal network is underlying the P300 of the present study.

The N200 is consistent with the visual awareness negativity [24,25] concept and the idea of posterior local recurrent activity [26]. Our N200 component occurs somewhat later and is less reliable than the usual N200 reported previously. This may be due to the faint stimulation. A similar explanation is offered by [18]. The facts showing that ERP correlates of correct perception have been found at a shorter latency range exemplified by N100-150 [27] can be explained as a result of the considerably higher contrast/intensity of the stimuli used, which leads to the speed-up of awareness-related processing and shorter latencies of the negative ERP components reflecting this.

We also did not observe early EEG components in the seen condition (e.g. N100) for the present paradigm. Again, it is likely that these signals are too faint and/or unreliable for the low contrast stimuli used in the present study. This interpretation is backed up by another study [18] where weak stimulation was used. The resulting very small post-stimulus brain response at 100 ms did not differ between conditions. Thus, the present results confirm that such early responses do not seem to be markers of direct conscious perception.

Taken together, our findings show that if a set of heterogeneous stimuli is used, whose identify cannot be predicted by the subject, the two widely reported correlates of consciousness – the N200 and P300 – are reliably observed. However, the onset latencies of these components still showed large variability. Importantly, part of this variability can be attributed to the particular set of trials selected for the condition without conscious perception. These results indicate that any conclusions about the NCC onset timing that are based on data from a single study with its specific stimuli and procedure, are likely to be misleading.

## ACKNOWLEDMENTS

We thank Annika Tallinn for her help during the development of the paradigm and the collection of data. We also thank Dr. Lucia Melloni for her helpful advice during the preparation of the manuscript.

## REFERENCES

1. Sergent C, Baillet S, Dehaene S. Timing of the brain events underlying access to consciousness during the attentional blink. Nat Neurosci. 2005;8(10):1391-1400.

2. Del Cul A, Baillet S, Dehaene S. Brain dynamics underlying the nonlinear threshold for access to consciousness. PLoS Biol. 2007;5(10):e260.

3. Koivisto M, Revonsuo A. Event-related brain potential correlates of visual awareness. Neurosci Biobehav Rev. 2010;34(6):922-934.

4. Pins D, Ffytche D. The neural correlates of conscious vision. Cereb Cortex. 2003;13(5):461-474.

5. Aru J, Bachmann T. Boosting up gamma-band oscillations leaves target-stimulus in masking out of awareness: Explaining an apparent paradox. Neurosci Lett. 2009;450(3):351-355.

6. Aru J, Bachmann T, Singer W, Melloni L. Distilling the neural correlates of consciousness. Neurosci Biobehav Rev. 2012;36(2):737-746.

7. De Graaf TA, Hsieh PJ, Sack AT. The 'correlates' in neural correlates of consciousness. Neurosci Biobehav Rev. 2012;36(1):191-197.

8. Navajas J, Ahmadi M, Quian Quiroga R. Uncovering the mechanisms of conscious face perception: a single-trial study of the n170 responses. J Neurosci. 2013;33(4):1337-1343.

9. Melloni L, Schwiedrzik CM, Müller N, Rodriguez E, Singer W. Expectations change the signatures and timing of electrophysiological correlates of perceptual awareness. J Neurosci. 2011;31(4):1386-1396.

10. Schurger A, Sarigiannidis I, Naccache L, Sitt JD, Dehaene S. Cortical activity is more stable when sensory stimuli are consciously perceived. Proc Natl Acad Sci U S

A. 2015;112(16):E2083–E2092.

11. Luck SJ. An Introduction to the Event-Related Potential Technique. Cambridge (MA): MIT Press; 2005.

12. Maris E, Oostenveld R. Nonparametric statistical testing of EEG- and MEG-data. J Neurosci Methods. 2007;164(1):177-190.

13. Picton TW, Bentin S, Berg P, Donchin E, Hillyard SA, et al. Guidelines for using human event-related potentials to study cognition: Recording standards and publication criteria. Psychophysiology. 2000;37:127–152.

14. Quian Quiroga R. Obtaining single stimulus evoked potentials with wavelet denoising. Physica D. 2000;145:278–292.

15. Dehaene S, Changeux JP. Experimental and theoretical approaches to conscious processing. Neuron. 2011;70(2):200-227.

16. Bachmann T. Microgenetic approach to the conscious mind. Amsterdam: John Benjamins; 2000.

17. Koivisto M, Revonsuo A. The role of selective attention in visual awareness of stimulus features: electrophysiological studies. Cogn Affect Behav Neurosci. 2008;8(2):195-210.

18. Sekar K, Findley WM, Poeppel D, Llinás RR. Cortical response tracking the conscious experience of threshold duration visual stimuli indicates visual perception is all or none. Proc Natl Acad Sci U S A. 2013;110(14):5642-5647.

19. Polich J. Updating P300: An Integrative Theory of P3a and P3b. Clin Neurophysiol. 2007;118(10):2128–2148.

20. Muller-Gass A, Macdonald M, Schröger E, Sculthorpe L, Campbell K. Evidence for the auditory P3a reflecting an automatic process: elicitation during highly-focused continuous visual attention. Brain Res. 2007;1170:71–78.

21. Ojanen V, Revonsuo A, Sams M. Visual awareness of low-contrast stimuli is reflected in event-related brain potentials. Psychophysiology. 2003;40(2):192-197.

22. Dehaene S, Kerszberg M, Changeux JP. A neuronal model of a global workspace in effortful cognitive tasks. Proc Natl Acad Sci U S A. 1998;95:14529–14534.

23. Dehaene S, Sergent C, Changeux JP. A neuronal network model linking subjective reports and objective physiological data during conscious perception. Proc Natl Acad Sci U S A. 2003;100:8520–8525.

24. Koivisto M, Revonsuo A. An ERP study of change detection, change blindness, and visual awareness. Psychophysiology. 2003;40(3):423–429.

25. Wilenius-Emet M, Revonsuo A, Ojanen V. An electrophysiological correlate of human visual awareness. Neurosci Lett. 2004;354(1):38–41.

26. Lamme VA, Roelfsema PR. The distinct modes of vision offered by feedforward and recurrent processing. Trends Neurosci. 2000;23(11):571-579.

27. Bachmann T. Psychophysiology of visual masking: the fine structure of conscious experience. Commack (N.Y.): Nova; 1994.

# SUPPORTING INFORMATION

## S1 Text. Pre-experiment.

Not all of the 70 stimuli have the same threshold contrast. An earlier pilot experiment indicated that there are 5 groups of stimuli with roughly similar threshold contrasts within each group: 1) text; 2) solid graphical figures; 3) line-drawings of graphical figures; 4) solid forms of all other figures and 5) line-drawings of all other figures. The appropriate contrasts for these 5 groups of stimuli were determined with the help of a short pre-experiment prior to the main experiment. The pre-experiment was very similar to the main experiment (see *"Task and design"*), except that a separate set of stimuli including all the 10 stimulus types (see "Stimuli") was used. Each stimulus (19 in total) was presented twice on 4 adjacent contrast levels. The specific contrast levels were different for each of the 5 contrast groups. They were typical threshold contrasts (i.e. lead to 50% seen responses) for these groups as indicated by an earlier pilot experiment. Subjects had to report whether they perceived a stimulus on each trial. Based on the detection rates of the pre-experiment, individual threshold contrasts for each of the 5 groups of stimuli were estimated by the experimenter. Occasionally, some of the contrasts had to be readjusted after the first block of the main experiment if detection rate was lower than 25% or higher than 75% for a particular group of stimuli.

## S2 Text. Luminance values.

Stimuli were presented on a light gray background with a luminance of 51.6 $cd/m^2$. The luminance of the stimuli was 48.5 $cd/m^2$ on average (median = 49.5 $cd/m^2$, SD = 1.25 $cd/m^2$, range = 46.5 – 51 $cd/m^2$). The size of the stimuli was approximately 2.5 degrees of visual angle. Prior to the stimulus a fixation cross was presented. The size of the fixation cross was 0.35 degrees of visual angle and its luminance was 11.4 $cd/m^2$. The response screen contained the question "Did you see something?" in the Estonian language. The contrast of the text was also low (luminance of 24 $cd/m^2$), in order not to disturb the adaption of the eyes for very low contrast stimuli.

# S3 Text. Neighboring electrodes and cluster formation.

For each of the 60 electrodes its neighboring i.e. surrounding electrodes are defined within a fixed radius. Note, however, that the radius varies depending on cap size (S =3.5 cm, M = 3.7 cm, L = 4 cm). As a consequence, each electrode has maximally 4 nearest neighbors, one in each cardinal direction. But lateral electrodes have less neighbors of course. For example, 'Cz' has neighbors 'FCz', 'C1', 'C2' and 'CPz', but 'Iz' only has 'Oz' as a neighbor.

This neighborhood structure is used by the cluster permutation test algorithm to group together data points that exhibit significant differences between conditions IF these data points are neighbors in time AND in space (i.e. they occur on neighboring electrodes). Figure A shows one such cluster. It is the P300 component that exhibited significant differences between the seen and the unseen condition. Note that, although not all rows are next to each other in the two-dimensional electrode-by-time plot, the corresponding electrodes are actually neighbors on the EEG cap. They were therefore clustered together.

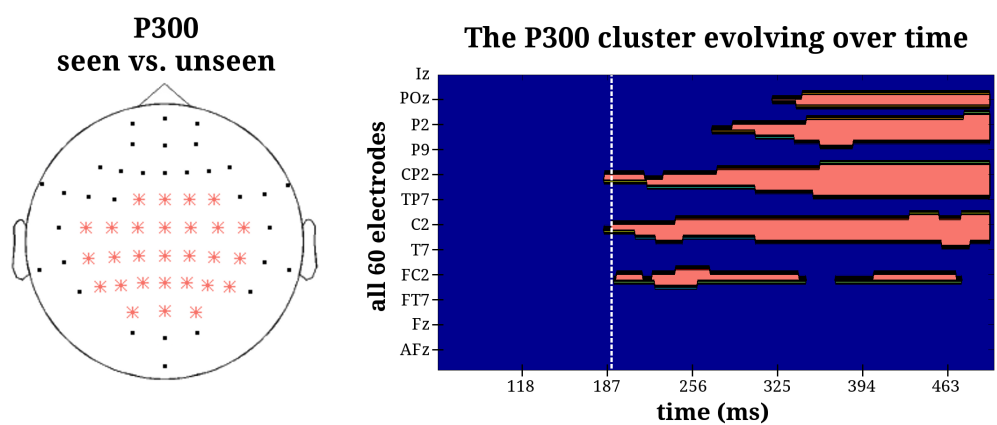

**Figure A. Typical result of a cluster permutation test.** On the right hand side is the electrode-by-time representation of the significant cluster (i.e. P300). All data points (electrode-time pairs) where the seen and the unseen condition exhibited significant differences are colored pink. Together they form the P300 cluster. The onset latency of this cluster is marked with the vertical dashed white line at 190 ms. This is the first time point where at least 4 neighboring electrodes show significant differences between conditions. On the left hand side are all electrodes belonging to the P300 cluster.

# S4 Text. Behavioral analysis.

As contrasts had to be readjusted occasionally during the main experiment (see S1 Text), detection rate also varied in accordance with the different levels of contrast. In order to eliminate this accountable variance from the behavioral results only those contrast levels are considered which comprise the most trials. Thus, 93.3% of all available trials are considered (SD over subjects = 9%; SD over types of stimuli = 2.8%). Results are comparable, however, when all available trials are considered.

## S5 Text. Correlation tests with ERP parameters.

In addition to the 12 correlation tests described in "Correlation tests" four extra correlation test were carried out between averaged ERP parameters and cluster onset latencies. For these tests denoised single trial data was first averaged for each electrode per condition. Then, peak amplitude and peak latency of N200/P300 (depending on the electrode) was noted for the seen and the unseen condition. Finally, these values were averaged over electrodes and over subjects and correlation tests were carried out with the onset latencies of the respective clusters. All the p-values (n = 16) were corrected for multiple comparisons with the Holm-Bonferroni method.

## S6 Text. Results for the correlation tests with ERP parameters.

The correlation between mean peak amplitude of the averaged seen trials and the P300 onset times was not significant (r = -0.26, t = -2.63, p = 0.089), but the correlation for mean peak amplitude of the averaged unseen trials was again highly significant (r = 0.53, t = 6.17, p = 2.5e-07). Similarly, the correlation of mean peak amplitude with N200 onset times for the seen trials was only marginally significant (r = 0.28, t = 2.9, p = 0.049). The same correlation for unseen trials was again highly significant (r = -0.47, t = -5.14, p = 1.9e-05). Figure B illustrates the results for these correlation tests.

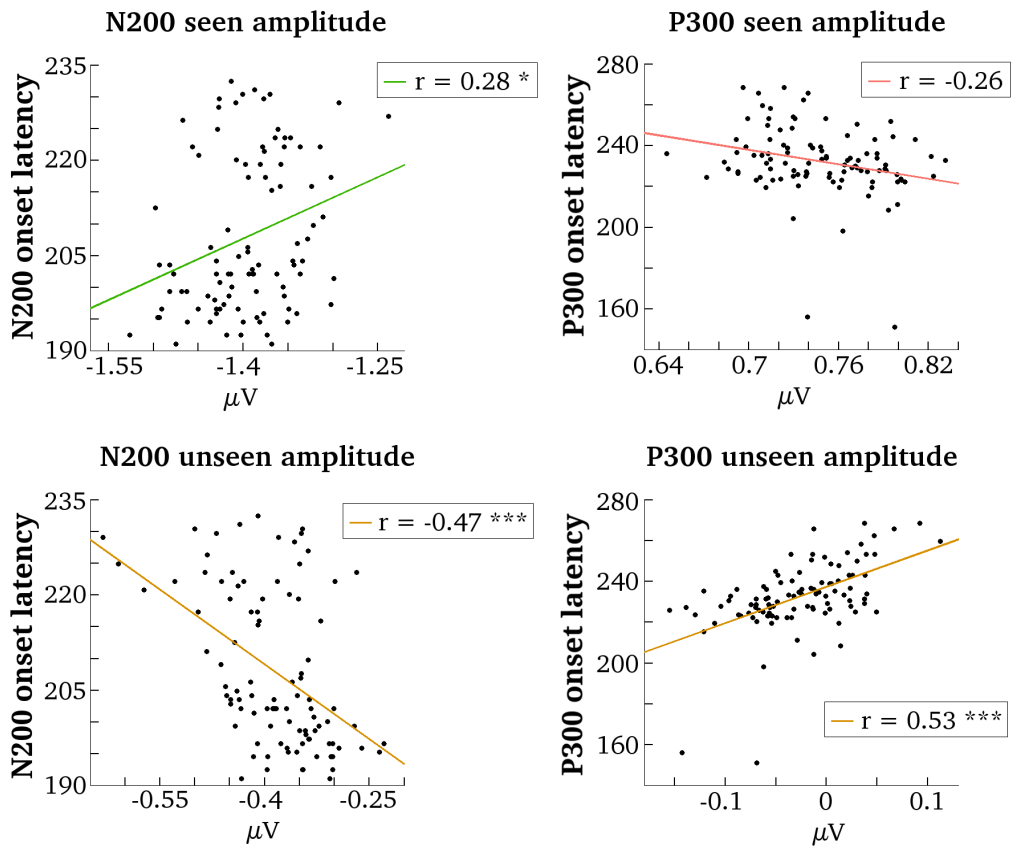

**Figure B.** Correlations between grand averages of N200 and P300 amplitudes (after

averaging the single trial responses per subject) and the respective gmNCC onset times (indicated in ms on the y-axes). Correlation tests are carried out separately for seen and unseen trials. P-values < 0.05 are indicated with *. P-values < 0.001 are indicated with ***.

## S7 Text. Detection rates for all stimulus types separately.

1. graphical line-drawings (m = 0.57, SD = 0.26)
2. graphical solid forms (m = 0.39, SD = 0.29)
3. short words (m = 0.87, SD = 0.19)
4. line-drawings of man-made objects (m = 0.5, SD = 0.2)
5. solid forms of man-made objects (m = 0.47, SD = 0.37)
6. line-drawings of faces (m = 0.49, SD = 0.2)
7. line-drawings of animated nature (m = 0.49, SD = 0.2)
8. solid forms of animated nature (m = 0.48, SD = 0.36)
9. line-drawings of inanimate nature (m = 0.36, SD = 0.2)
10. solid forms of inanimate nature (m = 0.47, SD = 0.38)

It is evident that the percentage of successfully perceived stimuli varies considerably between different stimulus types and even between single exemplars within a stimulus type. This, however, is not a problem for our present study. We are interested in the general markers of conscious visual perception. Such markers should not be affected by stimulus content variability. On the contrary, variance between stimuli can only strengthen any conclusions drawn from the results.

Furthermore, an ANOVA with factors stimulus type and conscious perception did not reveal any systematic effects on the proportion of trial numbers (main effect for stimulus type: $F(8,136) < 1.0$; main effect for conscious perception: $F(1,17) < 1.0$; interaction: $F(8,136) = 1.9$, $p = 0.07$). Note that for this ANOVA the stimulus type "short words" was excluded because it is already known that for this type detection rate is much higher than 50% on average.

## S8 Text. Late negativity.

The late negativity constituted a significant cluster on fronto-temporal electrodes. Like the P300, this negative cluster was significant on all 100 iterations and comprised of 21 electrodes on average (median = 21, SD = 0.4, range = 20 - 22). The mean onset latency of statistical significance for this negative cluster was 307 ms (median = 309, SD = 11 ms, range = 283 - 334 ms). Fig. C contains a histogram of the distribution. Again, this cluster was always significant until the end of the tested time period.

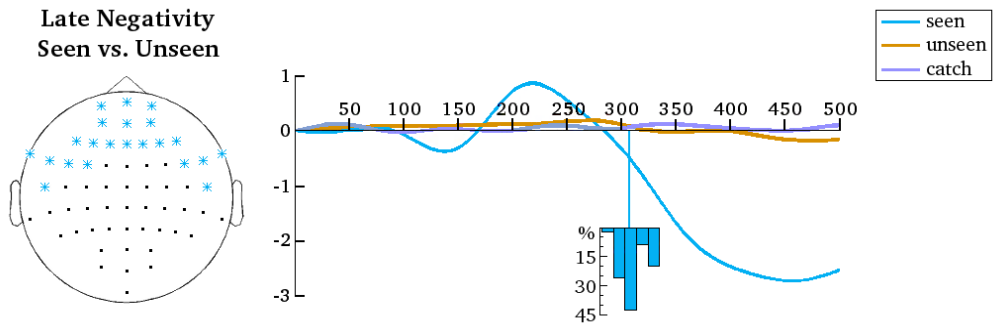

**Figure C. Summarized results for the late negativity.** ERPs are averaged over the indicated electrodes (left). These are all the electrodes that belonged to the respective cluster for at least 1 of the 100 contrastive analyses between the seen and the unseen condition. The histogram depicts the distribution of cluster onset times over the 100 contrastive analyses. Note that the distribution aligns with the time axis (in ms).

## S9 Text. Comparisons between the seen condition and baseline.

For comparisons between the seen condition with the baseline of the unseen condition post-stimulus data of the unseen condition was replaced by pre-stimulus data from the same trials. The same electrodes as for the main analysis were used. To remove any potential real components like CNV the baseline data was first detrended. Then all analysis steps were repeated just like before. The data was denoised and all the 100 contrastive analysis were carried out on the same sets of matched trials. Correlation tests with gmNCC onset latencies were conducted for mean peak amplitude, mean peak latency and standard deviation of peak latency from the seen trials and from the unseen trials (i.e. baseline activity) in the new time windows of observed variability in gmNCC onset latencies. The presently reported p-values are also corrected for 16 comparisons to make them comparable to the main results, but the significant changes in correlations with mean peak amplitude from the baseline activity are already visible when only the correlation coefficients are considered.

P300 was again significant on 100% of the iterations and always included all 9 electrodes that were selected for denoising. The new mean onset latency of P300 was 205 ms (median = 206 ms, SD = 8 ms, range = 180 - 219 ms) and P300 again always remained significant until the end of the tested time period. Compared to the results in the main text, variability in P300 onset latencies was now best correlated with the mean peak amplitude of the seen trials ($r = -0.3$, $t = 3.14$, $p = 0.03$). The correlation with mean peak amplitude of the "unseen" trials was not significant any more ($r = 0.2$, $t = 2.03$, $p = 0.54$). The onset times of P300 did not correlate significantly with mean peak latency for the seen nor for the "unseen" trials ($r = -0.15$, $t = -1.47$, $p = 1.0$ and $r = 0.13$, $t = 1.33$, $p = 1.0$, respectively). The correlations with mean latency variance were also not significant ($r = 0.13$, $t = 1.25$, $p = 1.0$ for seen trials; $r = -0.08$, $t = -0.77$, $p = 1.0$ for "unseen" trials). Figure D contains histograms of the respective distributions and correlations for mean peak amplitude.

N200 was also significant on 100% of the iterations and always included all 10 electrodes that were selected for denoising. Thus, compared to the results in the main text N200 is more reliably correlated with the seen condition. The new mean onset latency of N200 was 191 ms (median = 190 ms, SD = 3 ms, range = 179 - 198 ms). The new mean offset latency was 331 ms (median = 330 ms, SD = 3 ms, range = 325 - 337 ms). Thus, the new mean duration of the N200 was 140 ms (median = 139 ms, SD = 4 ms , range = 130 – 154 ms). Importantly, variability in N200 onset latencies was now best correlated with the mean peak amplitude of the seen trials (r = 0.39, t = 4.14, p = 0.001). The correlation with mean peak amplitude of the "unseen" trials was not significant any more (r = -0.26, t = -2.64, p = 0.13). The onset times of N200 did not correlate significantly with mean peak latency for the seen nor for the "unseen" trials (r = -0.13, t = -1.25, p = 1.0 and r = -0.15, t = -1.44, p = 0.32, respectively). The correlations with mean latency variance were also not significant (r = 0.13, t = 1.29, p = 1.0 for seen trials; r = 0.01, t = 0.1, p = 1.0 for "unseen" trials). Figure D contains histograms of the respective distributions and correlations for mean peak amplitude.

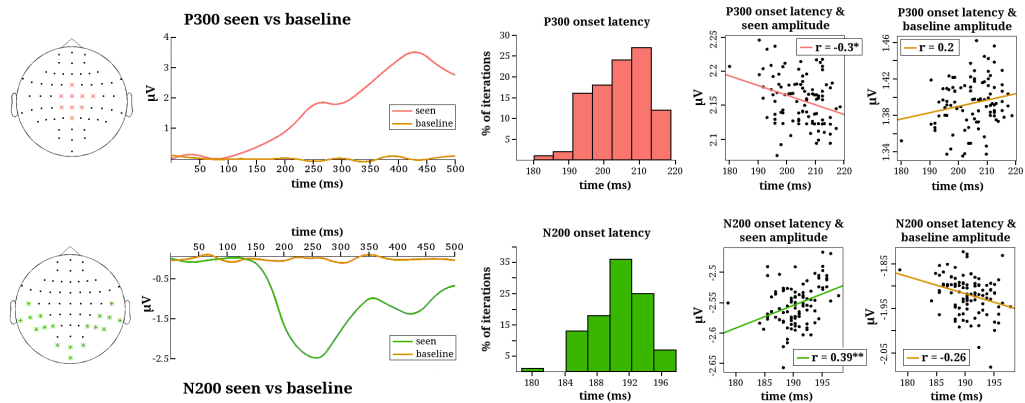

**Figure D. Results for the seen condition vs. baseline.** Denoised data is averaged over the indicated electrodes (left) and plotted over time for the seen condition and for baseline activity of the "unseen" condition (second from left). Histograms depict the distributions of gmNCC onset times over the 100 different contrastive analyses (third from left). The right hand side shows correlations between averages of single trial N200/P300 mean peak amplitudes and N200/P300 onset times (indicated in ms on the x-axes). Correlation tests are carried out separately for seen trials and baseline activity of the "unseen" trials. P-values < 0.05 are indicated with *. P-values < 0.01 are indicated with **.

### S10 Text. GmNCC for different stimulus classes and detection rates.

In order to investigate if certain stimulus characteristics have a reliable effect on N200/P300 parameters during the time period of their onset dependent samples t-tests comparing different groups of stimuli were carried out. First, only seen trials were considered in these analyses, because N200 and P300 seem to be uniquely associated with the seen condition. Second, all seen trials with extreme peak amplitude values during the time period of cluster onsets were removed from the data. Thus, all seen trials with peak amplitude values above or below two standard deviations from the grand average of peak amplitude were removed for the present analyses. Third, all seen text stimuli were removed for the present analyses, because stimuli of this type were detected more often than stimuli of other types. It is preferable to avoid this systematic unbalance between stimulus types.

One set of dependent samples t-tests was carried out to compare N200/P300 parameters for line-drawings and solid form images. The first group (i.e. line-drawings) consisted of all trials where line-drawings of graphical figures, man-made objects, animated nature, or inanimate nature were presented. The second group (i.e. solid forms) included all trials where solid form images of graphical figures, man-made objects, animated nature, or inanimate nature were presented. Thus, in addition to text stimuli, face stimuli were also excluded from this set of t-tests, because there was no solid form equivalent for face stimuli. On average, the were 114 trials available for line-drawings (median = 110, SD = 42.2, range = 45 - 228) and 101 trials for solid forms (median = 85, SD = 43.8, range = 38 - 164).

Another set of dependent samples t-tests was carried out to compare N200/P300 parameters for stimuli with higher and lower detection rates. The first group (i.e. high detection rate) consisted of those trials where a stimulus with > 50% detection rate was presented (each stimulus was presented 10 times and thus has a detection rate). The second group (i.e. low detection rate) consisted of those trials where a stimulus with <= 50% detection rate was presented. On average, the were 162 trials available for the high detection rate condition (median = 150, SD = 89.4, range = 40 - 367) and 83 trials for the low detection rate condition (median = 90, SD = 28.2, range = 31 – 131).

For both sets of t-tests mean peak amplitude, mean peak latency and the standard deviation of peak latency (for both the N200 and the P300) were compared between groups. Note that as for the correlation tests in "gmNCC onset variability explained by single trial parameters", these parameters were extracted from denoised single trials and averaged over the same representative electrodes of the respective clusters (see "Correlation tests" for more information). The results are presented in Table A below.

|  | **Peak Amplitude** | **Peak Latency** | **Peak Latency SD** |
|---|---|---|---|
|  | **T-tests for N200** | | |
| Stim. class | t = -2.0, p = 0.44, ges = 0.06 | t = -0.44, p = 1.0, ges = 0.003 | T = -1.98, p = 0.44, ges = 0.03 |
| Det. rate | t = 3.49, p = 0.03, ges = 0.11 | T = -1.15, p = 1.0, ges = 0.02 | T = 0.36, p = 1.0, ges = 0.002 |
|  | **T-tests for P300** | | |
| Stim. class | T = 2.87, p = 0.1, ges = 0.04 | T = 0.08, p = 1.0, ges = 0.0001 | T = -0.69, p = 1.0, ges = 0.009 |
| Det. rate | T = -3.38, p = 0.04, ges = 0.05 | T = -3.53, p = 0.03, ges = 0.16 | T = 2.5, p = 0.18, ges = 0.06 |

**Table A.** All t-tests have 17 degrees of freedom. All p-values are Holm-Bonferroni corrected for 12 tests in total. P-values < 0.05 are marked yellow. For effect size estimates we report generalized eta squared (ges).